\documentclass[sigconf]{acmart}
\usepackage{xspace}
\usepackage{enumitem}
\usepackage{circledsteps}
\usepackage{subcaption}
\usepackage{afterpage}

\usepackage{amsmath}
\usepackage{algorithm}
\usepackage[beginComment=//~,beginLComment=//~,endLComment=~,italicComments=true]{algpseudocodex}
\usepackage{listings}
\usepackage{graphicx}
\usepackage{flushend}
\usepackage{booktabs}
\usepackage[normalem]{ulem}
\usepackage{amsfonts}

\usepackage{tikz}
\usepackage{circledsteps}

\AtBeginDocument{%
  }

\renewcommand\footnotetextcopyrightpermission[1]

\def\regioncount{50\xspace}

\def\carbonunit{g$\cdot$CO$_2$eq/kWh\xspace}
\def\emissionunit{g$\cdot$CO$_2$eq\xspace}
\def\cotwo{CO$_2$\xspace}

\usepackage[many,theorems,skins,breakable]{tcolorbox}
\definecolor{tcbcolback}{RGB}{245,243,253}
\definecolor{tcbcolbox}{RGB}{254 218 173}
\definecolor{tcbcolframe}{RGB}{245,243,253}
\newtcolorbox{visionbox}[2][]{%
    colback=white!12,
    coltitle=black,
    colframe=teal!50,
    fonttitle=\bfseries,
    title=#2, 
    sharp corners,
    rounded corners=southeast,
    boxrule=0pt,
    enhanced,
    drop fuzzy shadow,
    #1, 
    top=3pt,bottom=2pt,left=3pt,right=3pt
    }

\begin{document}

\title{LiteCast: A Lightweight Forecaster for Carbon Optimizations}
\newcommand{\systemName}{\texttt{LiteCast}\xspace}

\author{Mathew Joseph}
\affiliation{%
  \institution{UC Santa Cruz}
  \country{USA}
}

\author{Tanush Savadi}
\authornote{Work performed during the Open Source Research Experience (OSRE) at UC Santa Cruz.}
\affiliation{%
  \institution{University of Massachusetts Amherst}
  \country{USA}
}

\author{Abel Souza}
\affiliation{%
  \institution{UC Santa Cruz}
  \country{USA}
}

\begin{abstract}
Over recent decades, electricity demand has experienced sustained growth, driven primarily by structural shifts such as the widespread electrification of transportation and the accelerated expansion of Artificial Intelligence (AI). Grids have managed the resulting surges by scaling generation capacity, incorporating additional resources such as solar and wind, and implementing demand-response mechanisms. Altogether, these policies influence a region's carbon intensity by affecting its energy mix: in some cases, they displace higher-emission sources during peak periods and lower emissions, while in others, upward variability derives from carbon-intensive peaking units that increase emissions. To mitigate the environmental impacts of consumption, carbon-aware optimizations often rely on long-horizon, high-accuracy forecasts of the grid's carbon intensity that typically use compute intensive models with extensive historical energy mix data. In addition to limiting scalability, accuracy improvements do not necessarily translate into proportional increases in savings. Highlighting the need for more efficient forecasting strategies, we argue that carbon forecasting solutions can achieve the majority of savings without requiring highly precise and complex predictions.  Instead, it is the preservation of the \textit{ranking} of forecasts relative to the ground-truth that drives realized savings. In this paper, we present \systemName, a lightweight time series forecasting method capable of quickly modeling a region's energy mix to estimate its carbon intensity.
\systemName requires only a few days of historical energy and weather data, delivering fast forecasts that can quickly adapt to sudden changes in the electrical grid. Our evaluation in \regioncount worldwide regions under various real-world workloads shows that \systemName outperforms state-of-the-art forecasters, delivering 20\% higher savings with near-optimal performance, achieving 97\% of the maximum attainable average savings, while remaining lightweight, efficient to run, and adaptive to new data.

\end{abstract}




\maketitle

\section{Introduction}

The last decades have witnessed global electricity consumption increase at an accelerated pace, driven by a combination of structural and technological shifts. 
Contributing factors include electrification of the residential~\cite{bayer2023modeling}, industrial, and transportation sectors, alongside the rapid proliferation of Artificial Intelligence (AI) applications and the large-scale expansion of data center infrastructures~\cite{iea-report}.
This surge has underscored the growing need to monitor and mitigate the environmental impacts of energy use.
Many of these initiatives have focused primarily on improving energy efficiency and implementing accounting offsets, in which certificates are swapped in exchange for the procurement of clean energy from solar and wind projects~\cite{maji2024green}.
While these policies have accelerated grid decarbonization, they also present new challenges: the variable and less predictable nature of renewable generation disrupts traditional power flows and system dynamics. 
As renewable integration deepens, ensuring grid reliability, controlling costs, and minimizing environmental impacts requires increasingly effective and coordinated, real-time planning.

In response, researchers have begun exploring alternative strategies to reduce the environmental impact and improve the carbon efficiency of electricity consumption. 
To achieve carbon-efficiency, both supply- and demand-side approaches have been adopted.
On the supply side, consumers use power purchase agreements (PPAs) of renewable energy sources, such as wind and solar, through long-term contracts to power their operations, commonly used in large-scale industries, e.g., data centers~\cite{maji2024green}. On the demand side, researchers have developed techniques to modulate demand and its associated emissions by shifting it to times and locations with low-carbon electricity. 
For instance, in buildings, flexible loads such as laundry, water heating, and residential electric vehicle (EV) can be deferred based on knowledge of low-intensity grid conditions. Likewise, ``spatial strategies'' can be employed to cloud workloads with significant geodistributed flexibility that allow them not only to change time, but also location to better align execution with the availability of low carbon energy~\cite{souza2023casper, gsteiger2024caribou, sukprasert2024limitations, google-co2-dc, goldverg2025towards}.

Generally, most carbon-aware optimizations assume near perfect knowledge of the grid's carbon intensity~\cite{wiesner2021let}, which may not always be accurate, limiting the potential of their solutions~\cite{grid_observability}. 
These optimizations have been made possible by the emergence of third-party services such as Electricity Maps~\cite{electricity-map} and WattTime~\cite{watttime} that provide real-time and predictive mechanisms with carbon and energy data, enabling the coordinated optimization of demand-side consumers and distributed compute workloads~\cite{electricitymap-forecast, wiesner2021let}.
State-of-the-art models apply complex, compute-intensive deep neural networks and ensemble models to understand the intra- and inter- spatiotemporal relationships between the various energy sources to forecast up to the next four days of a grid's carbon intensity~\cite{carboncast, yan2025ensembleci, zhang2023gnn}.
For instance, CarbonCast~\cite{carboncast} applies a two-tier neural network using historical energy, carbon, and weather data to improve carbon intensity forecasts.
Although these methods generally achieve high prediction accuracy across many regions---with few exceptions in grids with high wind penetration---their effectiveness in scheduling demand loads to optimally reduce their carbon footprint remains uncertain.
In addition, while these solutions are theoretically sound, many lack computational scalability, as they rely heavily on large volumes of historical data to perform well. 
This hampers low-latency performance, making real-time, adaptive operations more complex, time-consuming and costly in practice.
This is especially important in today’s evolving power grids, which undergo frequent upgrades and not only enhance inter-grid connections but also incorporate increasingly complex local energy systems, often featuring significant energy storage such as batteries.

In this paper, we introduce \systemName, a lightweight forecasting mechanism that enables fast, adaptive spatiotemporal carbon-optimizations.
In contrast to state-of-the-art solutions that prioritize accuracy through complex models, \systemName leverages the insight that carbon-aware scheduling does not always require precise carbon intensity data to minimize emissions. 
Instead, we argue that for effective scheduling, models should focus less on minimizing error metrics (such as MAPE, RMSE, or MSE), and more on accurately capturing the \textit{timing} and \textit{ordering} of carbon intensity forecasts because they better drive carbon-aware scheduling decisions to reflect real emissions. 
\systemName uses a SARIMAX method (Seasonal Auto Regressive Integrated Moving Average with Exogenous Regressor) that incorporates seasonality along with exogenous variables such as weather and demand forecasts across various worldwide regions. More importantly, \systemName requires less than 5\% of the historical data required by complex models, and is capable of being retrained in a very low-latency manner, without using large-scale resources. 
We experiment \systemName with various types of continuous and non-continuous loads and across various worldwide regions with varying power grid profiles. 
Compared to an oracle with perfect future knowledge, our approach achieves up to 97\% of the potential savings while remaining highly lightweight and efficient to run.
In evaluating our hypothesis, this paper makes the following contributions:
\begin{itemize}[leftmargin=0.4cm]
    \item \textbf{Lightweight Forecasting.} We present \systemName, a lightweight model that uses a standard time-series method along with exogenous data to understand and forecast the grid's carbon-intensity. 
    \item \textbf{Large-scale data analysis.} We analyze the performance of \systemName across \regioncount regions worldwide. 
    \item Together with \systemName, we introduce a heuristic that enhances forecasts by leveraging the temporal flexibility of jobs, dynamically shifting their schedules as new information becomes available while still meeting their SLO guarantees.
    \item \textbf{Large-scale analysis of forecasting implications on carbon-aware optimizations.}  We leverage state-of-the-art carbon-aware temporal workload shifting approaches, and spatial workload migration strategies, to quantify the actual carbon savings using \systemName. Our analysis  in \regioncount worldwide regions under various real-world workloads reveals that \systemName outperforms state-of-the-art forecasters, delivering 20\% higher savings with near-optimal performance, achieving 97\% of the maximum attainable average savings, while remaining lightweight, efficient to run, and adaptive to new data.
\end{itemize}
\label{sec:introduction}

\newcommand{\win}[2]{\sum_{k=0}^{L-1} #1_{#2+k}}

\section{Background}

In this section, we provide background on regional electricity grids, types of energy sources, 
carbon intensity signals, 
and carbon-aware optimizations.

\subsubsection*{Carbon Efficiency Signals} 
Carbon efficiency is measured in \emissionunit (grams equivalent of \cotwo) emitted per kilowatt-hour (\carbonunit), and varies over time and location, depending on the mix of energy sources supplying electricity to the grid~\cite{hitchin2002carbon}. Since fossil fuels have high carbon intensity and renewable have low or zero emissions, the carbon intensity of grid electricity depends on the relative share of each source. 
Two common carbon intensity signals are used to quantify carbon emissions from electricity consumption~\cite{sukprasert2024implications}: the average and marginal signals.
The average carbon intensity is estimated as the weighted average of the carbon emissions factors for all generators that satisfy current demand.
In the marginal paradigm, the carbon emission rate for the generator that responds to demand is used as the carbon intensity signal. As not all generators serve the added demand, only the \textit{marginal} generator does, and the higher or lower marginal emissions are assigned to consumers whose demand triggered it.
%
Although not fully capturing the exact impact of added demand, the average intensity signal is stable and observable, which is why the GHG protocol only requires reporting it~\cite{ghg-protocol}.
Similarly to other papers~\cite{carboncast, yan2025ensembleci, sukprasert2024implications}, our analysis focuses on the grid energy's \emph{average carbon-intensity}, which emissions fall under Scope 2 in the GHG protocol~\cite{ghg-protocol}.

\begin{figure}[t]
	\begin{minipage}{\linewidth}
		\includegraphics[width=\linewidth]{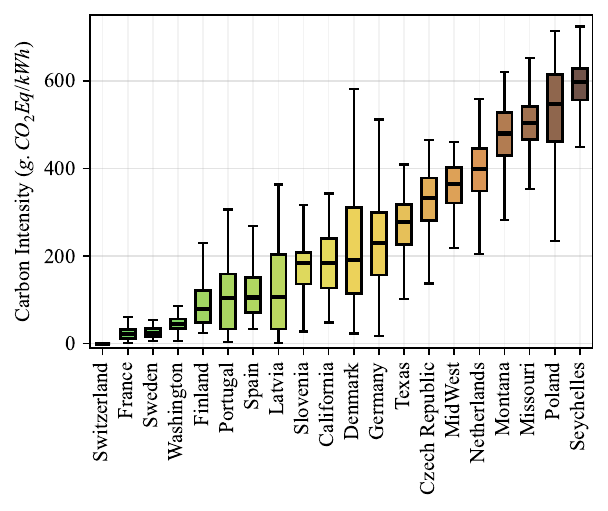}
        \vspace{-0.1cm}
	\end{minipage}\hfill
	\vspace{-0.5cm}
	\caption{\emph{The carbon intensity of energy supplied by the electric grid depends on the grid's energy mix and can vary by 3$\times$ and 10$\times$, temporally and spatially respectively.}}   
	\vspace{-0.5cm}
	\label{fig:intro-figures}
\end{figure}

\subsubsection*{Grid's Carbon Intensity} 
As discussed, a grid's carbon intensity fluctuates over time and depend on the mix of generators engaged to meet electricity demand. 
Demand follows human behavior (e.g., day vs. night, weekdays vs. weekends, holidays) and weather conditions, which affect heating and cooling needs. 
Each region is managed by a balancing authority tasked with meeting local demand using generators that differ in fuel type, capacity, ramp rate, and carbon intensity.
The electric grid uses 
a mix of generation sources, including fossil fuels (e.g., coal and natural gas), renewable options (e.g., hydro, wind, and solar), and low-carbon sources like nuclear and batteries. 
Renewable sources such as wind and solar are intermittent, impacting the variability of the generation mix.
The amount of energy produced and the percentage of energy being renewable will depend on the region. 

To illustrate, Figure \ref{fig:intro-figures} shows a grid's average carbon intensity collected from transmission system operators (TSOs) for several worldwide regions for 2023. The bar plots are sorted by yearly average intensities (lower regions to the left). The figure shows high magnitude variations, where temporal variations can reach up to 4$\times$  (e.g., in Denmark), and more than 700$\times$ across regions (spatial). 
The magnitude and variability of the carbon intensity of a grid depend on the mix of energy sources.
Notably, the average carbon intensity in Denmark can fluctuate by 3$\times$ (between 20 to nearly 600 \carbonunit), while in the Midwest (USA) it fluctuates by 2.15$\times$ (210 to 450 \carbonunit), and in California it has lower variations of 1.7$\times$ (from 219 to 370 \carbonunit). 
Typically, the higher the availability of intermittent sources of energy in a region, the higher the variations in carbon intensities the grid sees. 
Iowa has a high carbon intensity of around 500\carbonunit because only 31\% of its energy production uses renewable, low-carbon sources. 
California, on the other hand, has a 1.5$\times$ lower average (300\carbonunit) with high variability, since $\sim$67\% of its energy production is comprised of renewable sources. In contrast, $\sim$70\% of Iowa's energy derives from burning fossil fuels, resulting in a high average carbon intensity with low variability. 

%

%
\subsubsection*{Load Flexibility}  Demand encompasses any load interacting with the grid, including electric vehicle (EV) charging, household consumption, and data center workloads. 
Larger loads require greater grid responsiveness to maintain balance, directly affecting carbon emissions. 
A region’s carbon intensity is determined by the generation mix used to satisfy demand: in some cases, higher demand can displace carbon-intensive sources and lower emissions, while in others, variability necessitates reliance on peaking units with higher emission profiles, thereby increasing overall emissions.
An electrical load is considered \textit{flexible} when it can adapt to the availability of low-carbon energy. Flexibility has two main dimensions:  

\subsubsection*{Temporal flexibility} is the ability of a load to be postponed, extended, or shortened. Such loads can be \textit{interruptible} (e.g., EV charging, which can be delayed or paused) or \textit{continuous} (e.g., aluminum smelting, which cannot be stopped once started).  In computing, these loads are known as batch workloads.
Batch workloads generally include jobs with some degree of "slack" and thus may be delayed, or interrupted, although not indefinitely. Cluster schedulers often exploit this slack by deferring batch jobs' start time, i.e., forcing them to wait in a queue or periodically interrupting their execution~\cite{burns2016borg, 2003slurm}. 

\subsubsection*{Spatial flexibility} is the ability of a load to shift between different geographical locations. Spatially flexible loads can also be \textit{interruptible} or \textit{continuous}. For instance, web services are spatially flexible but continuous, since they can migrate with minimal disruption yet must remain always available.  

\noindent Loads may exhibit temporal flexibility, spatial flexibility, or both.

\subsubsection*{Carbon Forecasting} 


Effective forecasting is crucial for demand-response and carbon-aware optimizations, as it enables informed decision-making, supports planning and risk management, and plays a central role in minimizing the environmental impact of electricity consumption~\cite{gowrisankaran2016intermittency}.
Despite research on renewable~\cite{zhang2023improving} and energy demand~\cite{elamin2018modeling} forecasting, only recently solutions have been proposed to carbon intensity forecasting, which must necessarily combine multiple estimations~\cite{carboncast, yan2025ensembleci, zhang2023gnn}, similarly to multivariate methods, by averaging individual source carbon intensities.
Most work on carbon-aware optimization assume perfect knowledge of the grid's carbon intensity, a condition rarely met in real-world settings. Emerging research on online optimization is beginning to address some of these limitations by learning and adapting to the dynamics of the grid in real time~\cite{lechowicz2025learning}. However, they are not suitable for long-duration loads that require extended forecast windows to achieve carbon reductions.

\section{Carbon Optimizations} 

A scheduler leverages historical data and grid forecasts to determine when and where carbon intensity is low in order to plan the execution of a job\footnote{The terms 'job' and 'load' are used synonymously}. 
Each job is characterized by a length $L$ (its runtime) and a slack $T$ (the maximum allowable delay). 
The scheduling window is defined as $L+T$, representing the period within which a job must both start and complete. 
A carbon-aware scheduler uses a forecast of carbon intensity over this window to select a start time that ensures the job can be completed within its length $L$. 
We first formalize scheduling for continuous jobs under temporal flexibility. 
In our formulation and in \systemName, the term \textit{oracle emissions} refers to the realized future carbon emissions; the oracle is used only for evaluation and not available to the scheduler during forecasting.  

\noindent\textbf{Continuous Job.}  
Let $\hat f_t$ denote the forecasted carbon intensity at time $t$, $f^\text{act}_t$ denote the realized (actual) carbon intensity, and $f^\star_t$ denote the oracle forecast. 
Let the slack length be $S$ hours, the job length be $L$ hours, and the window be $t = 0,\dots,S+L-1$. 
The feasible start hours are $\mathcal{H}=\{0,1,\dots,S-1\}$. 
The scheduler selects the start time $s_{\text{pred}}(\mathcal{J})$ that minimizes the predicted emissions over the window $W=L+T$:  
\begin{equation}\label{eq:s-pred-cont}
s_{\text{pred}} \;=\; \arg\min_{h\in\mathcal{H}} \; \win{\hat f}{h}.
\end{equation}
The actual emissions realized by this schedule are then
\begin{equation}\label{eq:E-pred-cont}
E_{\text{pred}} \;=\; \win{f^\text{act}}{s_{\text{pred}}}.
\end{equation}

Analogously, $f^\star$ can be used to compute $s_{\text{oracle}}$ and $E_{\text{oracle}}$, which represent the minimum possible emissions under perfect foresight. 

\noindent\textbf{Interruptible Job.}  
Let $\mathcal{S}$ denote the set of hours within the window. 
An interruptible job may be distributed across non-contiguous hours, provided the total scheduled length is $L$. 
The scheduler chooses the hours of lowest forecasted intensity:  
\begin{equation}\label{eq:A-pred-inter}
A_{\text{pred}}
\;=\;
\arg\min_{\substack{A \subseteq \mathcal{S}\\ |A|=L}}
\;\sum_{t \in A} \hat f_t .
\end{equation}
The realized emissions are then given by
\begin{equation}\label{eq:E-pred-inter}
E_{\text{pred}}
\;=\;
\sum_{t \in A_{\text{pred}}} f^\text{act}_t .
\end{equation}

\noindent In both formulations, $E_{\text{pred}}$ depends on $\hat f$ only indirectly: the forecast determines the schedule ($s_{\text{pred}}$ or $A_{\text{pred}}$), while the realized emissions are evaluated against $f^\text{act}$.
Finally, to quantify the effectiveness of scheduler, we measure the ratio
\[
\rho \;=\; \frac{E_{\text{pred}}}{E_{\text{oracle}}}.
\]

\noindent That is, as $\rho$ approaches $1.0$, the predicted emissions $E_{\text{pred}}$ converge toward the oracle emissions $E_{\text{oracle}}$, indicating a more optimal outcome with minimized realized emissions.

\subsection{Motivation}

The prevailing focus in carbon intensity forecasting has been to maximize pointwise accuracy, minimizing deviations from the actual emissions as measured by standard error metrics such as mean absolute percentage error (MAPE) or root mean square error (RMSE). However, for the purpose of scheduling jobs while minimizing emissions, we argue that MAPE and related accuracy measures are not necessarily the most effective indicators of forecast quality.  

\begin{figure}[t]
	\begin{minipage}{1.1\linewidth}
		\includegraphics[width=\linewidth]{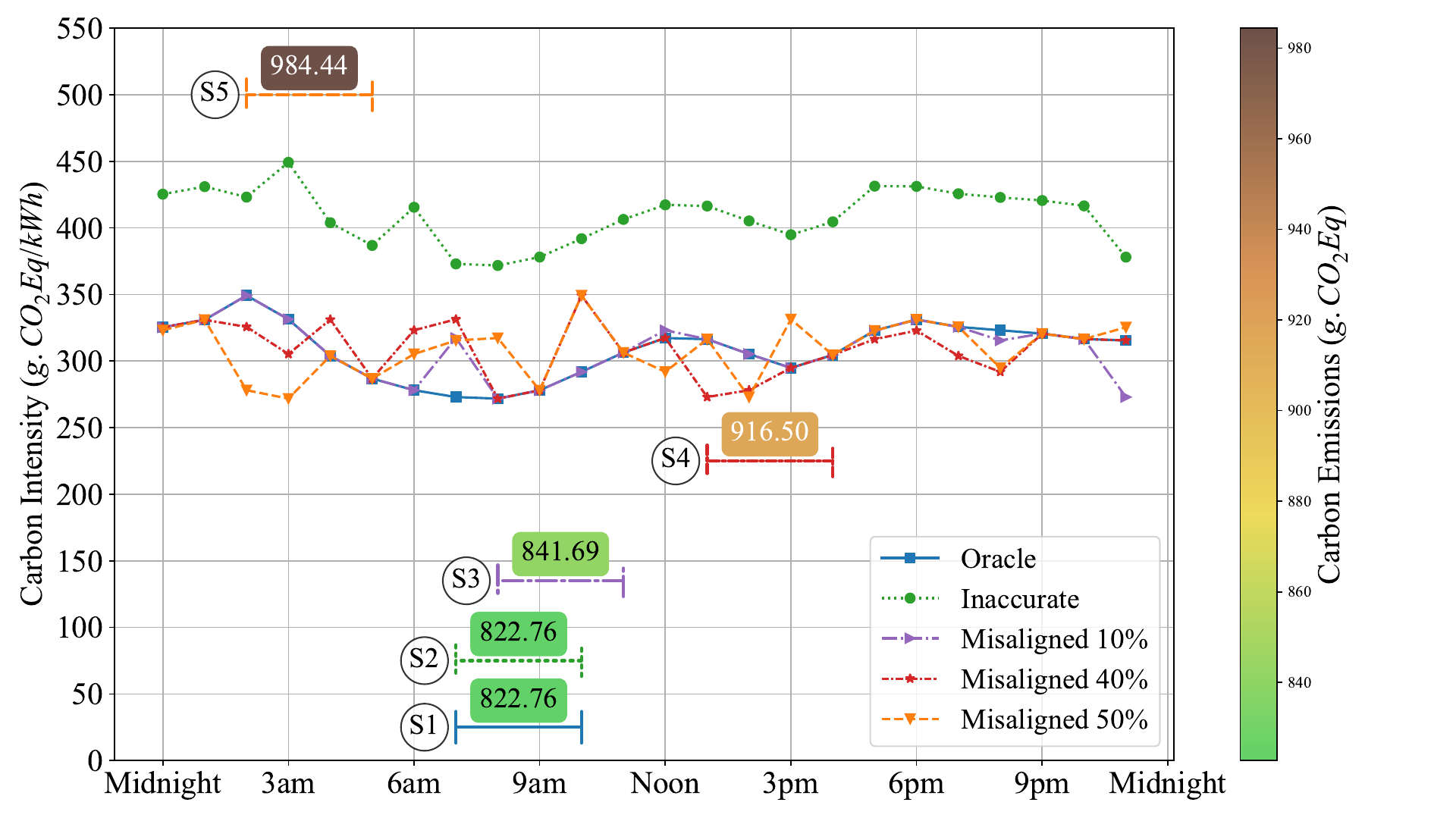}
        \vspace{-0.1cm}
	\end{minipage}\hfill
	\vspace{-0.3cm}
	\caption{\emph{Comparing Inaccuracy vs Misalignment in Texas over a 24 hour window with a 3 hour job}}   
	\vspace{-0.5cm}
	\label{fig:background-mape-v-concordance}
\end{figure}

Figure~\ref{fig:background-mape-v-concordance} illustrates this with a motivational example in which different forecasts are used to schedule a 3-hour continuous job on the ERCOT (Electric Reliability Council of Texas) grid. The figure compares four forecasts of varying error levels against an oracle forecast that has perfect future information about the grid. Using the oracle, the scheduler identifies the optimal schedule, \Circled{S1}, with 822.76~$g.\,CO_2Eq$. In contrast, forecasts with 10\%, 40\%, and 50\% misalignment relative to the oracle, exhibit low MAPE (1--5\% errors), yet lead to schedules \Circled{S3}--\Circled{S5} that perform increasingly worse in terms of emissions, up to 20\% higher than \Circled{S1}. By comparison, an ``inaccurate'' forecast with a high MAPE of 32\% error yields a schedule identical to the oracle’s \Circled{S2}, achieving the same optimal emissions.  
The critical difference lies in \textit{concordance}, which measures how well the ordering ranks of forecasted values match those of the oracle. While the misaligned forecasts achieve low MAPE, their concordance falls from 90\% to 50\%, leading to suboptimal schedules. The inaccurate forecast, despite its high MAPE, maintains a concordance of 95\%, enabling the scheduler to select the optimal schedule.  
This observation is supported analytically. In Equation~\ref{eq:s-pred-cont}, $\hat f$ aims to approximate $f^\text{act}$. Since realized emissions are separated from the forecast in Equation~\ref{eq:E-pred-cont}, it suffices that $\hat f$ correctly identifies the lowest-emission period at the same time as $f^\text{act}$. Under this condition, the same $s_{\text{pred}}$ and thus the optimal $E_{\text{pred}}$ will be chosen.  

We operationalize this insight with \systemName, a lightweight and adaptable system designed for seamless integration, enabling near-optimal carbon-aware load scheduling at both small and large scales scenarios.

\label{sec:background}
\section{\systemName Design}

This section describes \systemName, a lightweight grid's carbon-intensity forecaster and carbon-aware scheduler. The goal of our work is to optimize the operational carbon footprint of various types of workloads at various locations around the world when the expected grid energy intensity's information is not available.

\subsection{Overview}
\begin{figure}[t]
	\begin{minipage}{\linewidth}
        \includegraphics[width=1.1\linewidth]{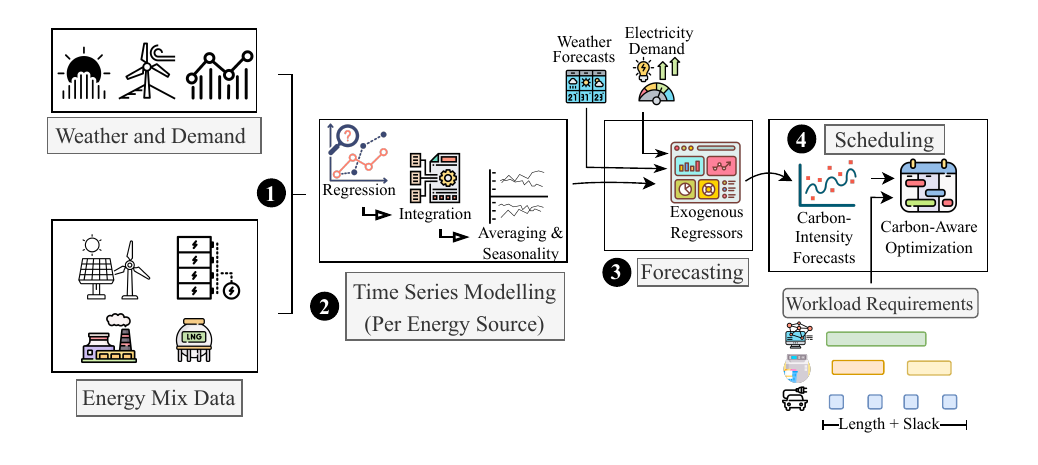}
        \vspace{-0.1cm}
	\end{minipage}\hfill
	\vspace{-0.5cm}
	\caption{\emph{\systemName Design.}}   
	\vspace{-0.5cm}
	\label{fig:litecast-arch}
\end{figure}

Figure \ref{fig:litecast-arch} depicts \systemName architecture. It includes four tandem components:

\noindent\CircledText{1} \textit{Energy and Weather Data}: This component aggregates time-series energy mix and weather data available from public sources such as the EIA~\cite{eia-monthly} and ENTSOE~\cite{entsoe-monthly}. It collects historical data on the regional electricity generation mix in a specific region (e.g., coal, gas, solar, etc.), their associated carbon emission factors (direct or lifecycle), weather forecasts~\cite{rda-data}, and the historical forecasts of electricity demand. LiteCast only requires seven days worth of historical data.

\noindent\CircledText{2} \textit{Modeling}: The energy and weather data are modeled using SARIMAX (Seasonal Auto Regressive Integrated Moving Average with Exogenous Regressor), a time series forecasting method that can incorporate exogenous (external) variables to enhance prediction accuracy. This module comprises three key components: (i) Regression (AR), which captures dependencies between current and past observations; (ii) Integration (I), which applies differential calculus to stabilize the time series by removing trends; and (iii) Averaging and Seasonality (MA), which accounts for patterns in prior forecast errors and incorporate seasonal variations, such as hourly or daily cycles, which is then forwarded to the Forecasting component.

\noindent\CircledText{3} \textit{Forecasting}: \systemName's forecasting module integrates external variables---including historical data and real-time forecasts of weather and electricity demand obtained from third-party services---to enhance SARIMAX-based predictions of the energy mix via its exogenous variable support. These forecasts are subsequently aggregated to yield hourly average carbon intensity estimates across varying time horizons.
For long-term horizons (e.g., $>24$ hours), historical data are combined with real-time forecasts to extend predictions to the required length. Such extended forecasts are particularly relevant for batch workloads with substantial temporal flexibility, such as AI training jobs with slack times of up to 168 hours. In these cases, \systemName produces multi-day forecasts (e.g., 7-day windows) that enable the scheduler to align workload execution with periods of lower carbon intensity.
When long-term forecasts are unavailable from external providers such as the EIA or ENTSOE, \systemName employs its regression module to generate the necessary forecasts for the exogenous variables specified by the user. 
\systemName can choose between SARIMAX and Prophet \cite{taylor2018forecasting}, depending on which one achieves a lower error in forecasting, when forecasting exogenous variables.

\noindent\CircledText{4} \textit{Scheduling}: Our carbon-aware scheduler uses the carbon-intensity forecasts, along with workload SLO (Service Level Objectives) requirements. Workloads often have flexible latency requirements and include jobs with some degree of ``slack'', allowing them to be delayed or paused intermittently, although not indefinitely. Our scheduler takes advantage of this flexibility by postponing job start times or periodically interrupting their execution to execute during the lowest carbon periods following the forecasts.

\subsection{Model}
The goal of our model is to predict the hourly average carbon intensity by forecasting the proportion of each source (e.g., solar, wind, gas, coal, etc.) used in electricity generation for the next hours and days, while using external factors to drive estimations.
Let $y$ model one time series for an energy source, and $\mathbf{x}$ be an exogenous vector influencing the forecast, but not part of the source itself, such as weather forecasts (e.g., solar irradiance, wind speed, temperature), energy demand forecast, hour of day, and day of the week.
The model forecast the value $y_t$ at time $t$ using external variables $x_t$ as follows:
\begin{equation}
\begin{aligned}
   y_t &= \phi_1 y_{t-1} + \Phi_1 y_{t-24} + \theta_1 \epsilon_{t-1} + \Theta_1 \epsilon_{t-24} \\
   &\quad - \Phi_1 y_{t-25} + \Theta_1 \epsilon_{t-25} + \boldsymbol{\beta}^\top \mathbf{x}_t + \epsilon_t .
  \label{SARIMAXEq}
\end{aligned}
\end{equation}

\noindent Equation \ref{SARIMAXEq} uses SARIMAX parameters defined by non-seasonal order $(p,d,q)=(1,0,1)$, and seasonal order $(P,D,Q)=(1,1,1)$ with a seasonal period $m=24$, representing hourly data with daily seasonality. The coefficients represent:
\begin{itemize}[leftmargin=0.3cm]
    \item $\phi_1$: non-seasonal autoregressive (AR) term with lag 1
    \item $\Phi_1$: seasonal AR term with lag 24 (hours)
    \item $\theta_1$: non-seasonal moving average (MA) term with lag 1
    \item $\Theta_1$: seasonal MA term with lag 24 (hours)
    \item $\epsilon_{t}$: noise error at time $t$
    \item $\boldsymbol{\beta}$: coefficient of exogenous input vector $x_t$
\end{itemize}
A SARIMAX model combines the strengths of standard ARIMA (Autoregressive Integrated Moving Average) with exogenous variables. 
It can model both past behavior and external factors, capture seasonality and trends, and unlike black-box models, it makes interpretable forecasts.
%










\noindent{\textbf{Ordering and Complexity}} 
SARIMAX effectively captures the concept of ``concordance''\footnote{\textit{Concordance Index} is a known metric in the biomedical research area of survival analysis quantifying the ability of a predictor to order samples~\cite{harrell1996multivariable}.}, which is the ability of a predictor to order samples by estimating the proportion of (forecasted) pairs correctly ordered among all comparable pairs in a dataset (ground-truth). 
When implementing spatiotemporal carbon optimizations, it is more important to capture the timely \textit{ordering} of the grid's carbon intensity than to accurately predict it. 
That is, simultaneously aligning forecasts with the grid's time and order, however inaccurate, effectively results in carbon-aware scheduling that follows the ground truth. This intuition is contrary to what most state-of-the-art methods implement~\cite{carboncast,yan2025ensembleci, zhang2023gnn}, which have a strong focus on optimizing accuracy metrics such as MAPE, RMSE, or MSE, and that do not consider forecasting ranking, in addition to require large historical data along with computational resources.
As we shall see, these properties result in a lightweight, low-latency forecasting engine, enabling \systemName's flexibility features that require dynamic, constant adaptation to real-time changes in grid conditions.

\begin{algorithm}[H]
\begin{algorithmic}[1]

\Require $J_{length}, J_{slack}, J_{arrival}, J_{start}$\\
\Comment{Job's length, slack, arrival and current starting times}
\Require $\{\mathcal{E}_t\}, {\hat f_t}$
\Comment{Energy-Weather and carbon intensities at time $t$}
\Require $\mathcal{CO}_2(\mathcal{J}, f), \mathcal{CI}(f)$
\Comment{$\mathcal{J}$'s emissions and concordance-index under $f$}

\Procedure{UpdateSchedule()}{}
\State $\mathcal{J} \gets \{J_{length}, J_{slack}, J_{arrival}, J_{start}\}$
\State $t \gets {\fontfamily{qcr}\selectfont time.now()}$
\If{$\mathcal{J}_{start} < t$}\\
    \Comment{Job not started, starting time may be updated}
    \For{$i \in \{[arrival, {arrival}+t]\}$}
            \State $\{\mathcal{E}_t\} \gets \{\mathcal{E}_{arrival}\}$ $\cup$ $\{\mathcal{E}_i\}$
            \Comment{Update training set with new data}
    \EndFor
    \State $\hat f_t \gets \systemName.train(\{\mathcal{E}_t\})$
    \Comment{Retrain \xspace\systemName with $\{\mathcal{E}_t\}$}
    \If{$\mathcal{CI}({\hat f_t}) > \mathcal{CI}({\hat f_{\mathcal{J}_{start}}})$}\\
    \Comment{New ${\hat f_t}$ has better concordance}
        \If{$\mathcal{CO}_{2}(\mathcal{J}, \hat f_t) < \mathcal{CO}_{2}(\mathcal{J}, \hat f_{\mathcal{J}_{arrival}})$}\\
        \Comment{$\hat f_t \xspace$ has new information to lower emissions}
            \State $new\_start \gets s_{pred}(\mathcal{J})$
            \State $\mathcal{J}_{start} \gets new\_start$
            \Comment{Update $\mathcal{J}$'s starting to $new\_start$, defined by Eqs. \ref{eq:s-pred-cont} or \ref{eq:A-pred-inter}}
            \State $\mathcal{J}_{arrival} \gets t$
            \Comment{Ensure job's arrival is updated to current $t$}
        \EndIf
    \EndIf
\EndIf
\\
\Return{\{$\mathcal{J}_{start}, \mathcal{C}_{\hat f_t}(\mathcal{J})$\}}
\EndProcedure
\end{algorithmic}
\caption{\systemName's Heuristic and Scheduling Logic}
\label{algo:scheduling_round}
\end{algorithm}

\subsection{Exploiting Job Flexibility}

The energy mix profiles across regions----caused by variability due to, e.g., weather, demand, and price variations---- have many implications on carbon forecasters, impacting the scheduling decisions and the resulting carbon emissions. 
With its simple structure, \systemName can be trained efficiently on recent historical data in a quick manner. 
Once trained, SARIMAX produces deterministic forecasts in the form of point estimates, without explicitly quantifying uncertainty, which assumes the historical patterns and relationships captured do not change in the future. This limitation is particularly relevant in the context of carbon optimizations in regions with highly volatile and weather-dependent energy systems.

\noindent\textbf{\systemName Heuristic. } Algorithm \ref{algo:scheduling_round} presents a dynamic heuristic that leverages \systemName's rapid retraining capability to adjust a job's initial schedule as new information becomes available. 
At each time step $t$, the algorithm first verifies whether the schedule of a job $\mathcal{J}$, originally planned to begin at $\mathcal{J}{start}$, can still be modified; this is valid when $t \leq \mathcal{J}{start}$ (Line 4). 
The updated weather and energy data ${\mathcal{E}t}$, collected since the job’s arrival $\mathcal{J}{arrival}$, are merged with the original dataset ${\mathcal{E}_{arrival}}$ (Lines 6–7). 
Since $\hat f_t$ incorporates both observed data (collected post-arrival) and forecasts, the combined dataset enables \systemName to retrain efficiently, producing a refined forecast model $\hat f_t$ (Line 8).
The heuristic evaluates this new model by comparing its concordance index $CI(\hat f_t)$ with that of the initial model $CI(\hat f_{arrival})$ (Line 9). 
If concordance improves, the algorithm further checks whether the expected emissions for a new schedule of $\mathcal{J}$---characterized by $J_{length}$, $J_{slack}$, and $J_{arrival}$---are lower than those of the original schedule (Line 10). 
When both conditions are met, the job's schedule is updated: \systemName replaces $\mathcal{J}{start}$ and $\mathcal{J}{arrival}$ with new values derived from $\hat f_t$ (Lines 11–13), which are then returned to the callee.

\label{sec:design}

\section{Evaluation setup and methodology}
\label{sec:methodology}

The main objective of our analysis is to evaluate and compare the effectiveness of \systemName and state-of-the-art predictors in exploiting spatiotemporal carbon-aware scheduling across different regions of the world to reduce emissions. We assess a predictor's performance against an \textit{actual} an ``oracle'' with perfect knowledge of the grid's true average carbon intensity. We quantify how each policy's carbon reductions differ depending on the use case and workload flexibility.
Finally, we also include the feasibility of running \systemName's heuristic (see Algorithm \ref{algo:scheduling_round}) that exploits the use of new, recent information to update scheduling decisions. 
This feature is essential to adapt to new grid's information that is available before scheduling decisions are made.


\subsubsection*{\textbf{Energy Traces.}} We use publicly available 2023 energy mix traces and weather data for \regioncount different geographical regions in the USA and Europe using EIA~\cite{eia-monthly}, ENTSOE~\cite{entsoe-monthly} and public weather data web APIs~\cite{rda-data}.
These traces report the hourly breakdown of the total of each energy source (depending on location), the total electricity demand (measured in kilowatt-hours, kWh), and weather forecasts (e.g., temperature, humidity, wind speed) from the Data for Atmospheric and Ocean Sciences Research Portal~\cite{rda-data} for all regions. 
The carbon intensity is measured in grams of carbon dioxide equivalent per kilowatt-hour (\carbonunit), in hourly granularity. The hourly granularity is the highest granularity for average carbon intensity data currently available from public sources. Although some regions observe large variations through a year (see \S\ref{sec:background}), the grid's carbon intensity rarely varies significantly within a 2-3 hour period, and thus higher granularity data would not significantly change the results of our analysis. 
The \textit{average} carbon intensity is calculated as the weighted average of the carbon emissions factors for all generators, and approach similar to previous works~\cite{carboncast, yan2025ensembleci, zhang2023gnn} (see \S\ref{sec:background}).
In addition, to weather data, we also use demand traces with forecasts of electricity consumption in a region.
Along with weather, demand forecasts are used as exogeneous variables in \systemName and help balance model predictions.
To compute the resulting carbon emissions, we attribute an average carbon factor for each source. Since carbon factors vary according to technology and regions, averaged values are used. We apply carbon factor values calculated by the IPCC (Intergovernmental Panel on Climate Change) for each energy source~\cite{bruckner2014technology}. 
Carbon intensities are reported for \textit{direct emissions}, defined as the operational emissions when an energy source is converted into electricity, and are used to account for scope 2 level emissions~\cite{scope1and2}. Finally, the resulting carbon emissions are calculated as the amount of carbon emitted into the atmosphere per unit of electricity generated by that source ($g/kWh$). 

\begin{figure*}[t]
    \begin{minipage}{0.24\linewidth}
        \includegraphics[width=\linewidth]{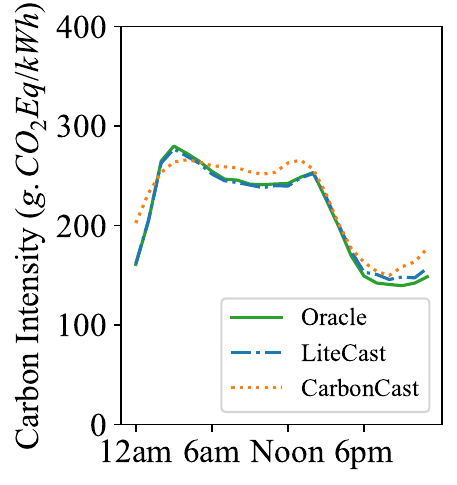}
        \subcaption{California}
    \end{minipage}
    \begin{minipage}{0.24\linewidth}
        \includegraphics[width=\linewidth]{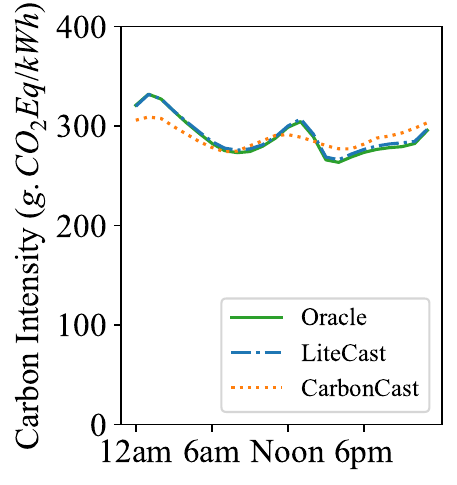}
        \subcaption{Texas}
    \end{minipage}
    \begin{minipage}{0.24\linewidth}
        \includegraphics[width=\linewidth]{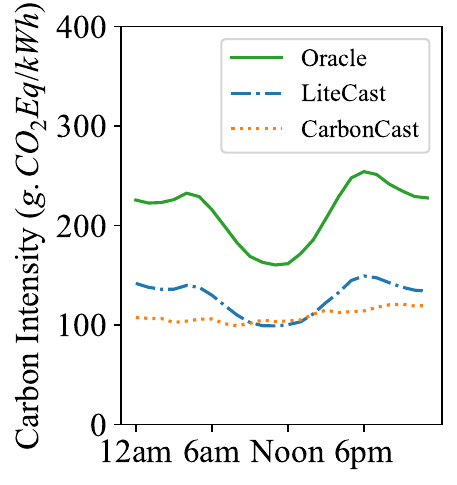}
        \subcaption{Denmark}
    \end{minipage}
    \begin{minipage}{0.24\linewidth}
        \includegraphics[width=\linewidth]{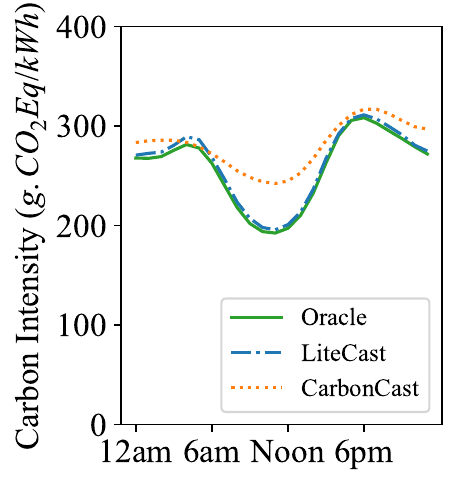}
        \subcaption{Germany}
    \end{minipage}
    \vspace{-0.4cm}
    \caption{\emph{Average carbon intensity (\%) 24H forecasts for US and European regions.}}  
    \vspace{-0.3 cm}
    \label{fig:avg_results}
\end{figure*}


\subsubsection*{\textbf{Workload and SLO}} Although our work applies to any flexible electrical load, we model an abstract load of length $L$ that consumes a unit power of $\sim$1 Watt and must be completed within a time horizon of $L + T$, where $T$ denotes the allowable slack representing a job's Service Level Objective (SLO). For all experiments and policies, $T$ is set between 12 and 168 hours, while the value of slack $L$ is specified per experiment and vary between 1 and 48H lengths.


\subsubsection*{\textbf{Spatial and temporal scheduling.}} 
Temporal flexibility includes loads with some degree of ``slack'' and thus may be delayed, or interrupted, although not indefinitely. We use two approaches: (i) continuous and (ii) interruptible executions. In the first approach, loads can be delayed but never paused, while in the second, jobs can be interrupted, delayed, and resumed later (checkpoint-restore). The forecasts of carbon intensity are used to indicate when the algorithm picks the $L$ lowest carbon slots within the $L+T$ horizon. 
For the spatial workload shifting, our policy migrates the job to the region with the lowest carbon intensity within all the possible geographical regions under analysis, inspired by approaches in prior work~\cite{souza2023casper, Strubell22}.

\begin{figure}[t]
    \begin{minipage}{0.24\linewidth}
        \includegraphics[width=\linewidth]{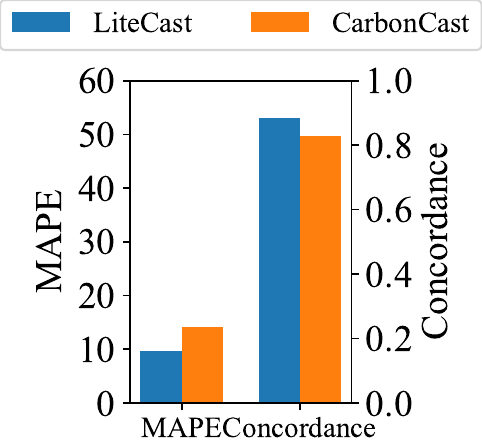}
        \subcaption{California}
    \end{minipage}
    \begin{minipage}{0.24\linewidth}
        \includegraphics[width=\linewidth]{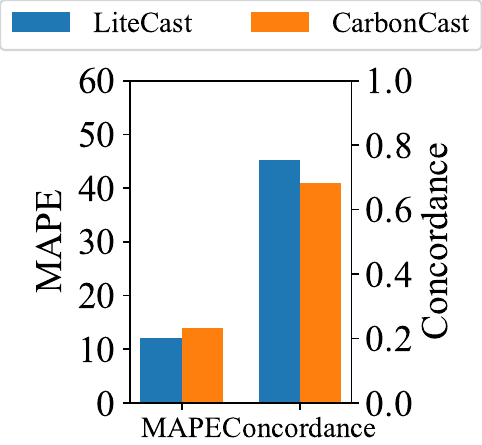}
        \subcaption{Texas}
    \end{minipage}
    \begin{minipage}{0.24\linewidth}
        \includegraphics[width=\linewidth]{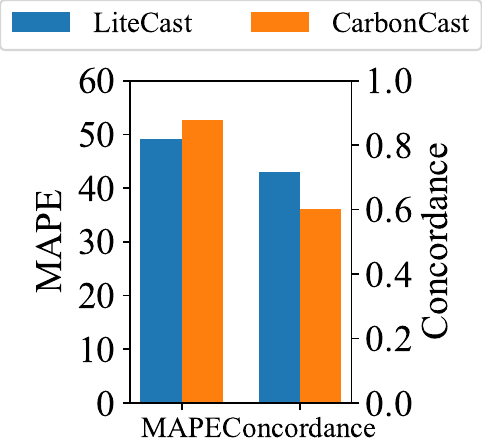}
        \subcaption{Denmark}
    \end{minipage}
    \begin{minipage}{0.24\linewidth}
        \includegraphics[width=\linewidth]{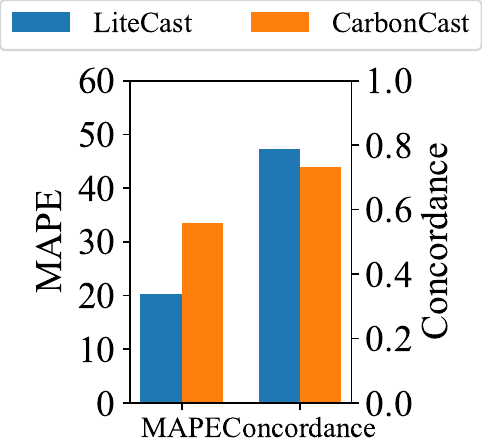}
        \subcaption{Germany}
    \end{minipage}
    \vspace{-0.4cm}
    \caption{\emph{MAPE and Concordance Index for US and EU regions.}}  
    \vspace{-0.5 cm}
    \label{fig:mape_production}
\end{figure}

\begin{figure*}[t]
    \begin{minipage}{0.33\linewidth}
        \includegraphics[width=\linewidth]{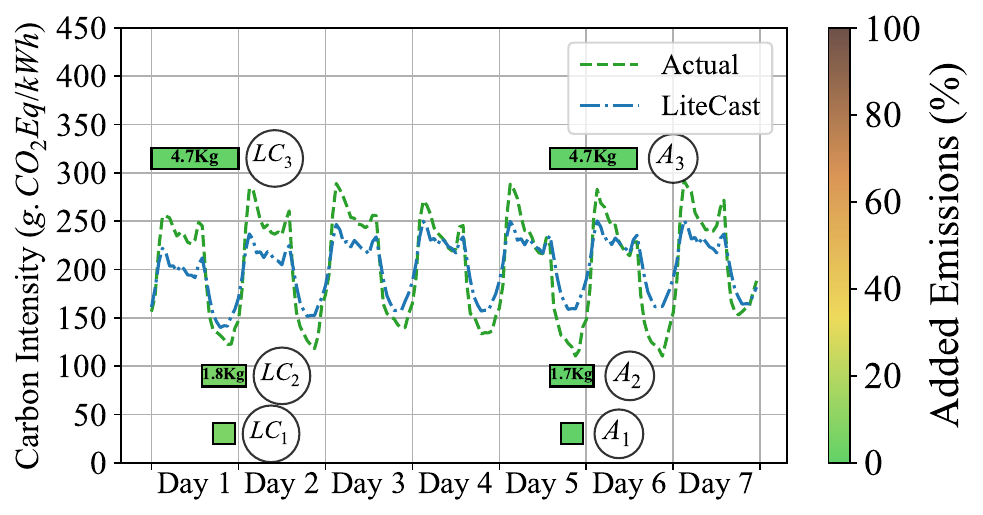}
        \subcaption{California}
    \end{minipage}
    \begin{minipage}{0.33\linewidth}
        \includegraphics[width=\linewidth]{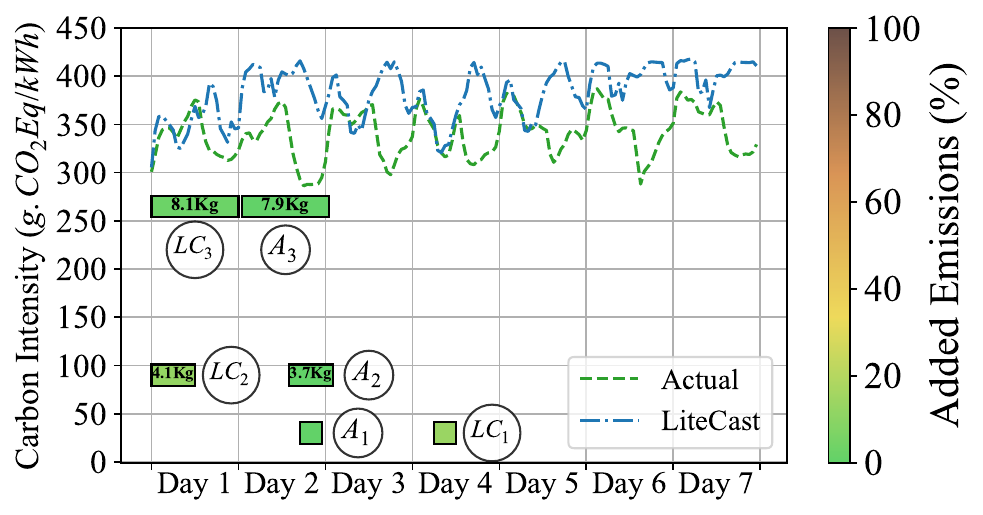}
        \subcaption{Texas}
    \end{minipage}\hfill
    \begin{minipage}{0.33\linewidth}
        \includegraphics[width=\linewidth]{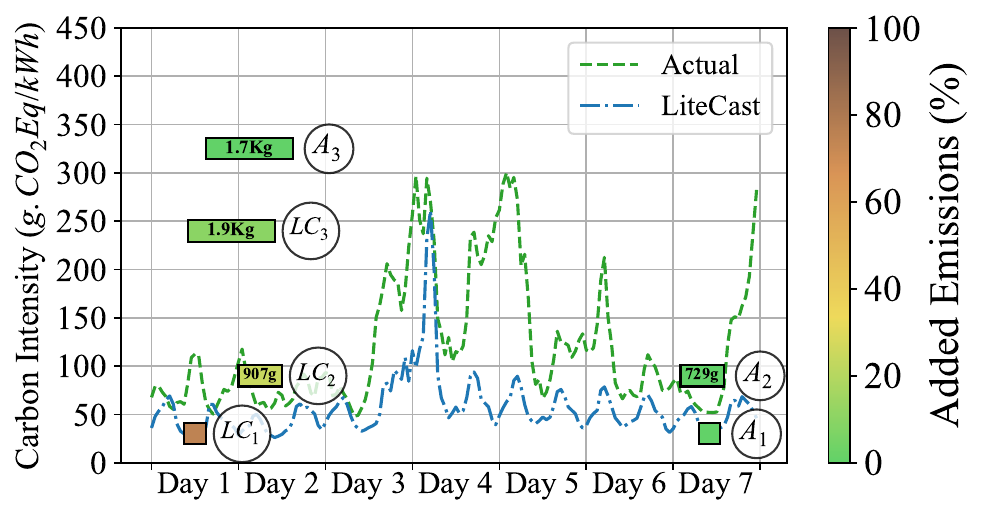}
        \subcaption{Denmark}
    \end{minipage}
    \vspace{-0.4cm}
    \caption{\emph{Forecast vs Actual sample schedules across three regions.}}  
    \vspace{-0.6 cm}
    \label{fig:temporal-samples}
\end{figure*}

\subsubsection*{\textbf{Emission Calculations.}} To evaluate the effectiveness of different predictors, we use three predictors to implement carbon-aware scheduling: (i) Persistence, (ii) CarbonCast, (iii) \systemName, and the (iv) Heuristic. The (i) Persistence method simply re-uses the last day's observations as predictions for the next day; (ii) CarbonCast utilizes a neural network and time-series transformer for its predictions, which is open-sourced and available. CarbonCast was originally designed to forecast up to 96H forecasts, but we have enabled it to support 168H forecasts. In addition, (iii) \systemName (see \S\ref{sec:methodology}), and (iv) uses \systemName along with our heuristic that updates job schedules following the availability of new data and prior to its start. 
For all methods, we first calculate the total \textit{direct} carbon emissions incurred from the generation of electricity. These emissions serve as comparative baselines and are attributed to a job that uses an ideal carbon-aware scheduler 
with perfect knowledge of the future carbon intensity, capable of selecting the most optimal time and location to execute a job, i.e., when and where the carbon intensity is the lowest possible while respecting the job length and slack constrains.
We next compute the carbon emissions associated with each scheduling policy, which uses their own forecasts to decide when and where a job is executed. 
The emissions associated with a job execution are attributed based on the specific, actual carbon intensity at the location and time period that the particular forecaster (i.e., Persistence, CarbonCast, \systemName, and Heuristic) chooses.
With perfect forecasts, the baseline and carbon-aware policy schedules align perfectly and differ otherwise, but are never lower than the actual baseline.
Finally, in order to compare different regions, we compute carbon emissions as percentages relative to the actual baseline.

\subsubsection*{\textbf{Setup.}} We implement \systemName and its heuristic in Python 3.10, using the default SARIMAX implementation from statsmodels~\cite{statsmodel}. The demand-forecasting module combines \texttt{statsmodels} and Prophet~\cite{prophet}, selecting the model with the lowest MAPE per region. All \regioncount+ experiments are executed on Nautilus~\cite{nrp} using a Kubernetes-based parallel cluster~\cite{burns2016borg}.
For a fair comparison, both CarbonCast and \systemName rely on similar grid and weather datasets. CarbonCast is implemented following the procedures described in its paper~\cite{maji2022carboncast} and public repository~\cite{carboncast}. Since CarbonCast requires historical data from past years to construct its model, we use 2022 energy and weather records to generate forecasts for 2023. In contrast, \systemName requires only seven days of historical weather and demand data from the job submission date.  While training sets larger than seven days can be configured to improve accuracy in certain regions, this comes at the cost of requiring longer historical records, which may not always be available. 
Furthermore, by incorporating exogenous variables, \systemName can adapt to grid conditions and, together with its heuristic, dynamically update forecasts under unforeseen scenarios.
This design makes \systemName more flexible to deploy and enables the use of an adaptive heuristic (Algorithm~\ref{algo:scheduling_round}) that exploits real-time data to refine scheduling decisions.

\section{Results}

In this section, we evaluate the performance of the four carbon optimizations described in Section \S\ref{sec:methodology}: Persistence and CarbonCast against \systemName and its heuristic.
In both temporal and spatial experiments, we consider jobs of various lengths and slacks, pinpointing the relative emissions savings across all \regioncount regions. 
We schedule jobs through all days of 2023, for job lengths and slacks of 1, 6, 12, 24, 48, 96, and 168H.

\subsection{\systemName Forecasts}

Figures \ref{fig:avg_results} and Figures \ref{fig:mape_production} show an hourly time series averaged over 2023 for two U.S. and two EU regions with high renewable penetration, along with their MAPE and concordance-index calculations for (a) California, (b) Texas, (c) Denmark, and (d) Germany.
As can be seen in the figure, compared to Oracle (continuous green lines), \systemName forecasts (blue lines) closely match the average trends in carbon intensity across the four regions.
In addition, Figures \ref{fig:mape_production}(a)--(d) show that \systemName outperforms CarbonCast in all regions, with an MAPE ranging from 9\% in California to 49\% in Denmark. In particular, although \systemName overestimates Denmark's wind production and lowering its average forecasts and incurring a large MAPE (Figure \ref{fig:mape_production}(c)), the model correctly detects the forecast sensitivity to changes with respect to the actual time series, outperforming CarbonCast's concordance index by 15\% (Figure \ref{fig:avg_results}(c).
Importantly, Figures \ref{fig:mape_production}(a)--(d) show \systemName's MAPE gains relative to CarbonCast are not proportional to the concordance index gains, indicating time-series model better capture the ordering ranking of forecasts.
In all regions, \systemName achieves 10\% higher concordance than CarbonCast, 
demonstrating that simpler time-series models can rival more complex models in practical scheduling scenarios.

\subsection{Implications on Carbon-Aware Scheduling}

The yearly average results are important to illustrate trends in the the overall performance across methods, but in reality, the actual hourly forecasts vary, impacting a method's resulting emissions. Figures \ref{fig:temporal-samples} present both actual and forecast samples for (a) California, (b) Texas, and (c) Denmark. We ignore Germany because its results were similar to those of Denmark. 
In temporal scheduling experiments, we want to compare the emissions of \systemName with the oracle (Actual) by submitting three different continuous job lengths of 3, 12, and 24H on Day 1, all with a slack of 168H.
Circles represent the selected schedules for each policy: \Circled{$A_i$} represent the oracle's schedules, while \Circled{$LT_i$} represent \systemName schedules, with index $i \in \{1:3H, 2:12H, 3:24H\}$, with rectangles representing the exact periods of time each job runs. The vertical heatbar represents the additional emissions as a percentage relative to the actual emissions, which are the greenest colors.
We can see that in California (Figure \ref{fig:temporal-samples}(a)), \systemName has very similar emissions to the actual, with a slight 5\% deviation for schedule \Circled{$LC_2$} when compared to \Circled{$A_2$}, despite being scheduled in different days, indicating a very precise hourly forecast ranking. 
Texas and Denmark are highly difficult regions due to the predominance of renewables, specifically wind (see Figure \ref{fig:intro-figures}), resulting in large swings in their hourly carbon intensity. Despite this, in Texas (Figure \ref{fig:temporal-samples}(b)), \systemName deviates only by 2\% for 24H lengths \Circled{$LC_3$}, and by 10\% for 12 and 3H lengths \Circled{$LC_2$} and \Circled{$LC_3$}. The same can be seen in Denmark (Figure \ref{fig:temporal-samples}(c)), with larger variations of 20\% for 12H jobs \Circled{$LC_3$} due to its even higher wind production variations.

\begin{figure}[t]
    \begin{minipage}{0.33\linewidth}
        \includegraphics[width=\linewidth]{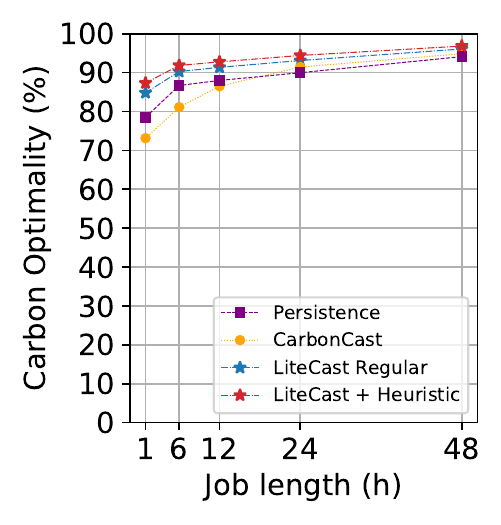}
        \subcaption{24H slack}
    \end{minipage}
    \begin{minipage}{0.33\linewidth}
        \includegraphics[width=\linewidth]{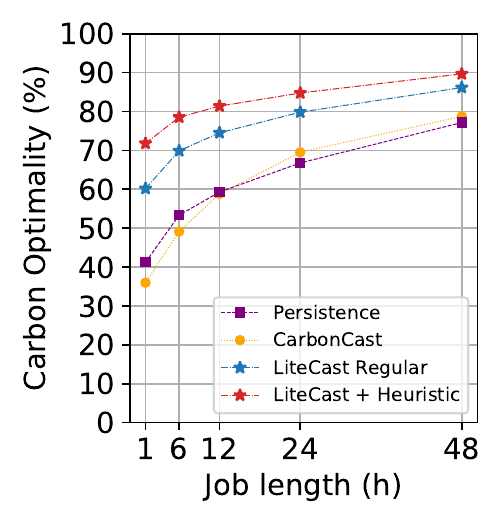}
        \subcaption{96H slack}
    \end{minipage}\hfill
    \begin{minipage}{0.33\linewidth}
        \includegraphics[width=\linewidth]{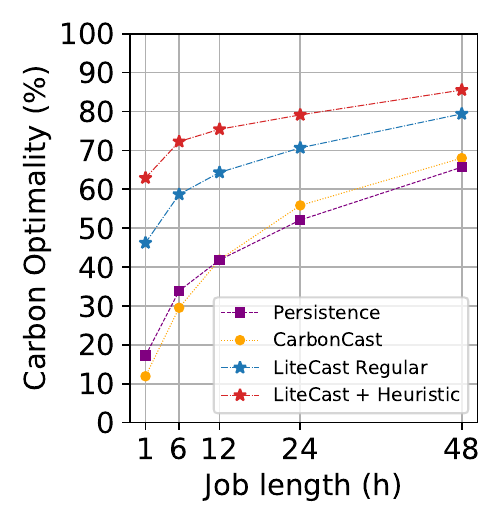}
        \subcaption{168H slack}
    \end{minipage}
        \begin{minipage}{0.33\linewidth}
        \includegraphics[width=\linewidth]{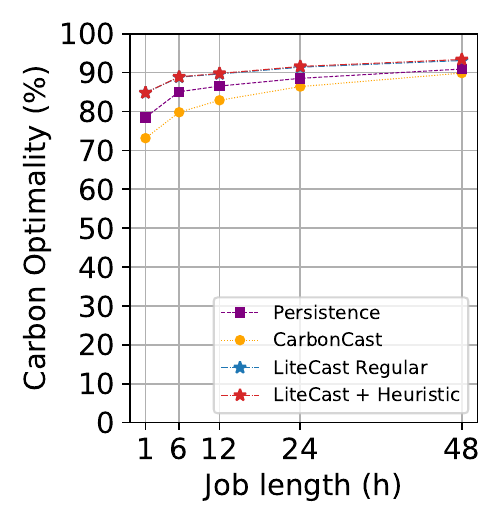}
        \subcaption{24H slack}
    \end{minipage}
    \begin{minipage}{0.33\linewidth}
        \includegraphics[width=\linewidth]{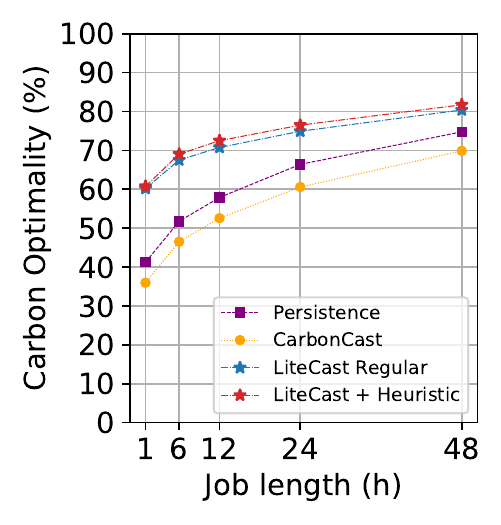}
        \subcaption{96H slack}
    \end{minipage}\hfill
    \begin{minipage}{0.33\linewidth}
        \includegraphics[width=\linewidth]{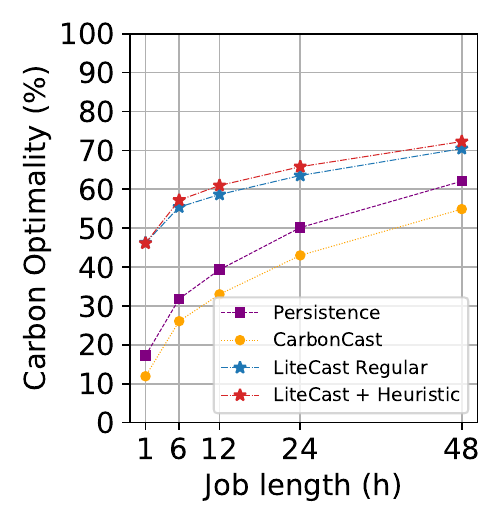}
        \subcaption{168H slack}
    \end{minipage}
    \vspace{-0.4cm}
    \caption{\emph{Average carbon optimality (\%) for continuous ((a)--(c)) and interruptible ((d)--(f)) jobs executions across all lengths and regions, with respect to the actual's emissions.}}  
    \vspace{-0.5 cm}
    \label{fig:temporal-continuous}
\end{figure}

Figures \ref{fig:temporal-continuous} present average results for both continuous and interruptable jobs in all \regioncount regions comparing all policies (i.e., \systemName and its Heuristic, CarbonCast, and Persistence) to actual emissions, that is, emissions resulting from schedules originating from an oracle with perfect knowledge. In these experiments, the carbon optimality metric quantifies the distance the emissions of each policy are from the lowest actual emissions.
The results for both continuous and interruptable jobs across all \regioncount regions are shown. 
Figures \ref{fig:temporal-continuous}(a)--(c) show the effect of larger forecast windows as the job length and the slack (24, 96, and to 168H) increases: the shorter the job length, the higher the added emissions, with \systemName achieving an average of $\sim$5-20\% lower emissions than CarbonCast, depending on the slack. \systemName + Heuristic in particular, can achieve up to 97\% carbon optimality on average for $< 24H$ job lengths, 90\% for $< 96H$ lengths, and 85\% for $< 168H$ lengths.
Differences between the heuristic and regular \systemName and CarbonCast
Figures \ref{fig:temporal-continuous}(d)--(f) show that for interruptible jobs that can be stopped and resumed, the effect of multiple hourly forecasts accentuates the increases in emissions with larger slacks. Unlike continuous jobs, interruptible jobs level-off at 5 and 10\% of added emissions for 12 and 48h slacks respectively.

\begin{figure}[t]
    \begin{minipage}{0.52\linewidth}
        \includegraphics[width=\linewidth]{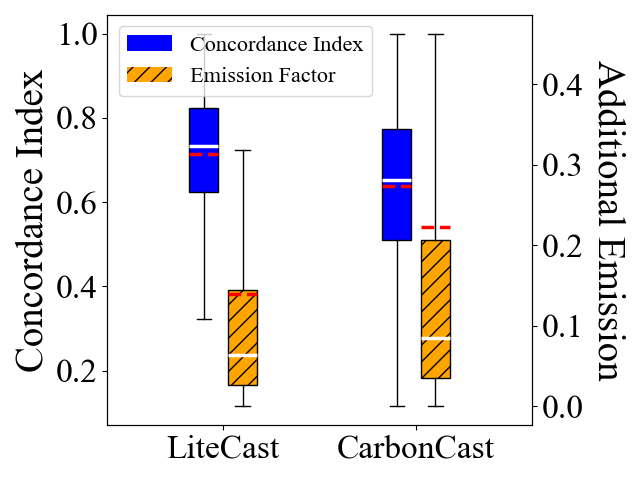}
        \subcaption{Overall Results}
    \end{minipage}\hfill
    \begin{minipage}{0.4\linewidth}
        \includegraphics[width=\linewidth]{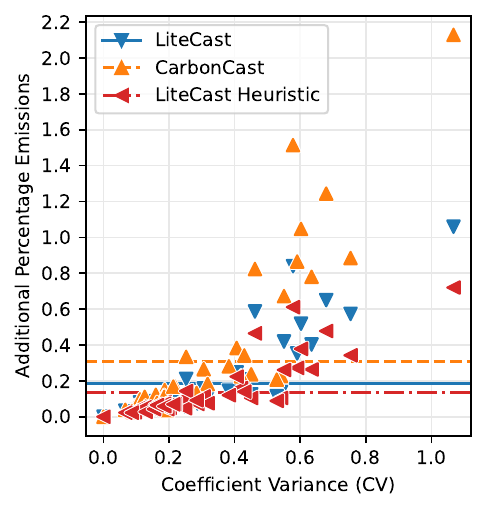}
        \vspace{-0.8cm}
        \subcaption{Regional Savings}
    \end{minipage}
    \vspace{-0.4cm}
    \caption{\emph{(a) Average concordance (\%) and additional carbon emissions (\%), and (b) overall regional savings comprising all jobs executions for all lengths and slacks, as a percentage with respect to the oracle's emissions.}}  
    \vspace{-0.7cm}
    \label{fig:temporal-takeaway}
\end{figure}

Finally, Figures \ref{fig:temporal-takeaway} present the (a) overall results for the concordance index (left y-axis), alongside the additional emissions (right y-axis), and a (b) scatterplot of the additional emissions for \systemName, CarbonCast, and the heuristic, all relative to the oracle, for all lengths of jobs, slacks, and across all regions, continuous and interruptible.
The x-axis in Figure \ref{fig:temporal-takeaway}(b) shows the Coefficient of Variation (CV), which calculates the ratio of the standard deviation ${\displaystyle \sigma }$ to the mean ${\displaystyle \mu }$ for each region, i.e., the higher the CV, the higher the variations of carbon intensity in a region.
The benefits of the Heuristic are noticeable, incurring only in ~17\% additional emissions across all \regioncount regions, down from 36\% for CarbonCast, and 5\% down from \systemName (regular, with no heuristic).
Figure~\ref{fig:temporal-takeaway}(a) demonstrates that a 15\% increase in concordance directly translates into an approximate twofold reduction in realized emissions, indicating that improvements in the correct ordering of forecasted carbon-intensity values enable the scheduler to more consistently select low-emission time slots.

\subsection{Cluster Scheduling}

\begin{figure}[t]
    \begin{minipage}{\linewidth}
        \includegraphics[width=\linewidth]{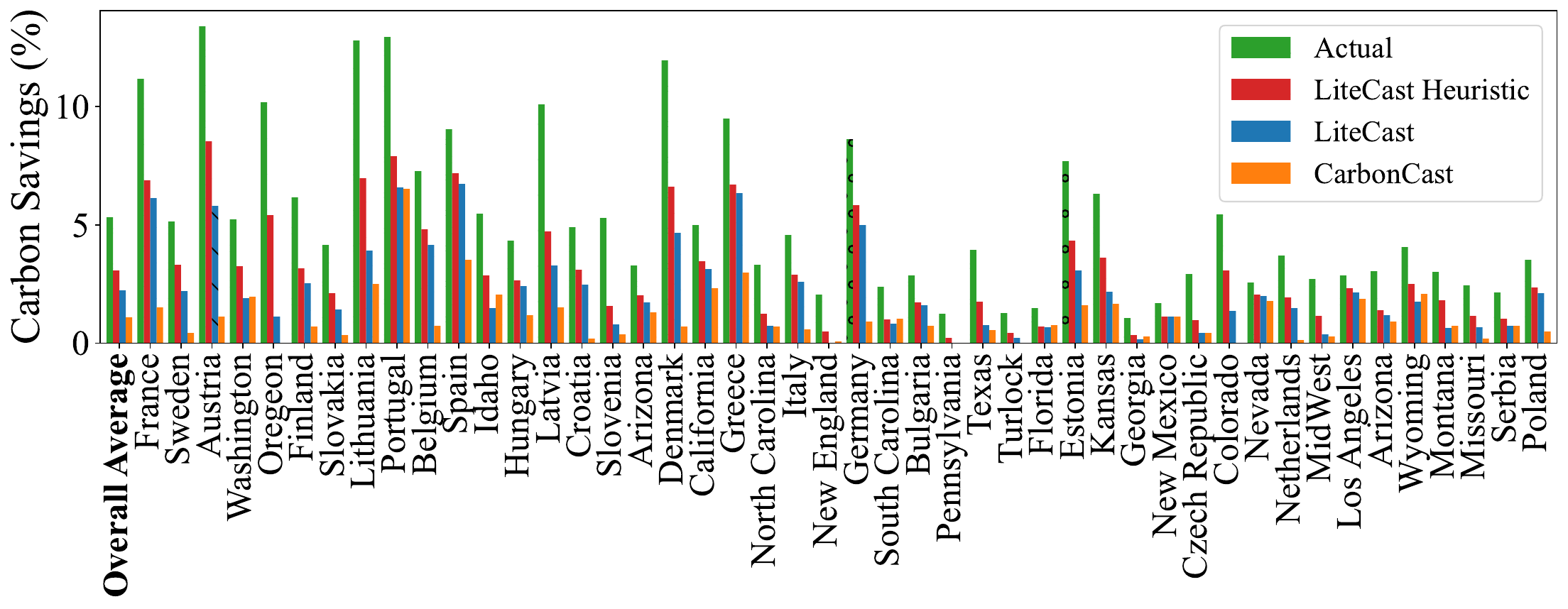}
    \end{minipage}
    \vspace{-0.4cm}
    \caption{\emph{Savings result comparisons (\%) for the KIT supercomputer workload across \regioncount regions.}}
    \vspace{-0.5cm}
    \label{fig:spatial-interruptible}
\end{figure}

Until this point, we have presented results for isolated load executions. In practice, however, carbon-aware schedulers must manage diverse cluster job types, each with varying lengths and slacks. To capture this heterogeneity, we evaluate using the KIT trace from the Parallel Workload Archive~\cite{kit}, which contains one and a half years of accounting records from the ForHLR II system at the Karlsruhe Institute of Technology in Germany. We replay the trace to simulate job submissions as if they were issued by real cluster users, assigning each job’s slack according to the waiting time recorded in the trace. Figure~\ref{fig:spatial-interruptible} reports the additional emissions for \systemName, its heuristic extension, CarbonCast, and the oracle-based Actual baseline. While \systemName achieves average savings of 3.4\% (approximately 2\% lower than the oracle), its heuristic consistently outperforms CarbonCast, yielding higher savings in 93\% of the evaluated \regioncount regions. The predominance of short jobs in the KIT trace~\cite{kit} naturally limits opportunities for optimization; nonetheless, the heuristic effectively exploits real-time information to refine forecasts and secure additional savings within short timeframes.

\begin{visionbox}{}
\noindent\textbf{Key Takeaway } \emph{
Compared to state-of-the-art forecasters, \systemName delivers up to 20\% higher savings while reaching 97\% of carbon-optimality. These results demonstrate that effective carbon-aware scheduling depends primarily on preserving forecast rankings, captured by metrics such as the concordance index, rather than maximizing precision.
}
\end{visionbox}

\label{sec:discussion}
\section{Discussion}

Our results highlight the trade-offs between lightweight statistical approaches such as SARIMAX, as implemented in \systemName, and more complex deep learning models such as CarbonCast. SARIMAX-based methods are inherently scalable for moderate-sized datasets and can accommodate diverse data frequencies with relatively low computational overhead. This efficiency makes them well suited for scenarios where rapid adaptation to evolving grid conditions is necessary, particularly when only limited historical data are available. At the same time, we acknowledge that scaling SARIMAX to very large datasets or high-frequency data streams remains challenging, especially in the presence of numerous exogenous variables or complex seasonal effects. As datasets and model structures grow, the computational complexity increases substantially, which may limit applicability in large-scale, real-time forecasting.
By contrast, deep learning models such as CarbonCast leverage parallel architectures to efficiently generate forecasts once trained, making them suitable for long-horizon predictions across large regions. However, the computational and memory requirements of their initial training phase remain substantial, restricting their scalability in practice. Moreover, unlike SARIMAX, deep learning methods often produce probabilistic rather than deterministic forecasts, which introduces uncertainty into the forecast ranking. Since our findings show that concordance—the preservation of the correct ordering of carbon-intensity forecasts relative to ground truth—is more consequential for realized savings than pointwise accuracy, this probabilistic nature can impact scheduling effectiveness, even when mean error metrics are favorable.
Importantly, our evaluation shows that \systemName achieves near-optimal performance (97\% of the maximum attainable savings) while remaining lightweight and adaptive. These results suggest that prioritizing forecast concordance over absolute accuracy yields more practical and scalable solutions for carbon-aware scheduling. 

\label{sec:related}
\section{Related Work}

\noindent\textbf{Carbon-Intensity Forecasters.}
CarbonCast is an open-source forecasting framework designed to estimate carbon intensity using publicly available datasets, such as those from the U.S. Energy Information Administration (EIA)~\cite{eia-monthly} and ENTSO-E~\cite{entsoe-monthly}, thereby eliminating dependence on third-party services like ElectricityMaps~\cite{electricitymap-forecast} or WattTime~\cite{watttime}. It delivers state-of-the-art hourly carbon intensity forecasts, extending up to 96 hours into the future.
More recently, EnsembleCI has been introduced~\cite{yan2025ensembleci}, which applies multiple ensemble models to identify the most suitable predictor for each region. While effective for improving forecast quality, EnsembleCI nor CarbonCast have yet been applied to carbon-aware scheduling, which is the purpose of this research.

\noindent\textbf{Carbon Optimizations.}
To exploit temporal fluctuations in carbon intensity, prior work has explored a range of scheduling and load-shifting strategies aimed at reducing both environmental impact and monetary costs~\cite{souza2023casper, sukprasert2024limitations}. For example, residential emissions can be reduced by scheduling flexible appliances—such as dishwashers or laundry machines—to run during off-peak or low-carbon periods~\cite{laicane2015reducing}. Similarly, energy storage systems, such as batteries, can be charged when electricity is cleaner or cheaper, and discharged during periods when the grid is dirtier or prices are higher. These strategies effectively shift demand away from carbon-intensive periods, aligning consumption with cleaner supply conditions, and all require energy forecasters to understand and plan for the near-term future.

\label{sec:results}
\section{Conclusion}

This work has demonstrated that carbon-aware scheduling does not require highly precise forecasts to unlock most of the achievable carbon savings. Instead, the critical factor is preserving the relative ordering of carbon-intensity predictions, ensuring that scheduling decisions align with the lowest-emission opportunities. To this end, we introduced \systemName, a lightweight and adaptive forecasting framework that leverages minimal historical data to generate fast, region-specific predictions. By design, \systemName reduces computational overhead while maintaining the flexibility to incorporate new information as grid conditions evolve. Our evaluation across \regioncount regions under diverse real-world workloads shows that \systemName consistently delivers near-optimal savings—achieving 97\% of the theoretical maximum—while outperforming state-of-the-art baselines by 20\%. These findings underscore that forecasting strategies should prioritize efficiency and concordance over accuracy alone, paving the way for scalable, carbon-aware optimizations in future power systems.
Future work may explore hybrid approaches, such as combining SARIMAX’s lightweight adaptability with ensemble methods or probabilistic adjustments from deep learning models, to balance efficiency with robustness across grid topologies and seasonal variations. 

\bibliographystyle{ACM-Reference-Format}
\bibliography{sample-base}






\end{document}